# Exploring Multi-Transition-Metal NASICON Frameworks as High-Performance Cathodes for Sodium-Ion Batteries

Santosh Behara[a,b], Achinthya Krishna Bheemaguli[b], Gopalakrishnan Sai Gautam*[b]

[a]Department of Metallurgical and Materials Engineering, National Institute of Technology Andhra Pradesh, Tadepalligudem 534101, India

[b]Department of Materials Engineering, Indian Institute of Science, Bengaluru 560012, India

E-mail: saigautamg@iisc.ac.in

**Abstract**

The search for sustainable, high-performance cathodes has driven a growing interest in sodium superionic conductor (NASICON)-type phosphate compounds for sodium-ion batteries (SIBs). To identify promising NASICON compositions containing earth-abundant transition metals (TMs) and to systematically examine the role of multiple TMs in influencing the various properties of NASICON cathodes, we employ density functional theory calculations in this work to investigate nine NASICON compositions containing Mn, Cr, and/or Fe, and spanning unary, binary, and ternary combinations. Our calculations reveal that unary systems, in terms of their Na intercalation phase behavior, exhibit well-defined stabilization at intermediate Na contents (x in $Na_xTM_2(PO_4)_3$), while binary and ternary systems display more complex phase behavior, with some systems showing a shift of thermodynamic minima from x = 3 to x = 2. Intercalation voltages highlight the dominant role of $Fe^{4+}/Fe^{3+}$ redox activity in elevating average voltages (~4.0 V), while Mn and Cr introduce intermediate-to-low voltage redox activity, respectively. Electronic structure data demonstrate non-systematic changes in the band gap, especially in systems containing multiple TMs. $Na^+$ mobility results identify mixed-TM frameworks as favorable, achieving $Na^+$ migration barriers in the 0.3-0.4 eV range. Importantly, we identify $Na_xMnFe_{0.5}Cr_{0.5}(PO_4)_3$ to be a promising ternary composition for subsequent experimental validation, offering an optimal intersection of phase stability, voltages, thermodynamic (meta)stability, and $Na^+$ migration barriers. Together, our study provides fundamental insights into the interplay between compositional complexity, thermodynamic stability, electronic structure, and ionic transport in NASICON cathodes, and offers actionable design principles for utilising multi-TM NASICONs as high performance cathode materials in SIBs.

**Keywords:** Sodium-ion batteries, NASICON cathodes, multi-transition-metal substitution, density functional theory, sustainability.

## 1. Introduction

The global transition toward renewable energy and electrified transportation has intensified the demand for advanced rechargeable battery technologies.[1,2] Lithium-ion batteries (LIBs) dominate the market owing to their high energy density, efficiency, and long cycle life.[3,4] However, concerns regarding the long-term sustainability of LIBs, arising from the uneven geographical distribution of lithium and key transition metals (TMs), have created economic, political, and supply chain vulnerabilities.[5–7] These challenges have accelerated the search for alternative energy storage systems,[8] among which sodium-ion batteries (SIBs) have emerged as a promising and cost-effective solution.[9,10] SIBs share many similarities with LIBs, including intercalation-based mechanisms and comparable design principles, making them attractive as "drop-in" replacement technologies that can utilize existing manufacturing infrastructure.[11,12] Nevertheless, their practical deployment critically depends on the development of high-performance electrode materials, particularly cathodes.[13–15]

The most widely studied SIB cathode classes are layered TM oxides (TMOs), Prussian blue analogues (PBAs), and polyanionic compounds.[16,17] Layered TMOs offer high specific capacities and facile synthesis but suffer from structural instability due to detrimental phase transitions during cycling.[18–20] PBAs exhibit favorable rate capabilities and moderate capacities but face intrinsic challenges such as cyanide-group decomposition and structural water, compromising safety and long-term stability.[21–23] In contrast, polyanionic frameworks,[15,24] particularly sodium superionic conductor (NASICON)-type materials, first reported by Hong and Goodenough,[25,26] have gained significant attention for their three-dimensional Na-ion ($Na^+$) diffusion pathways, high operating voltages, and robust open framework structures.[27–29] With the general formula $Na_xTM_2(XO_4)_3$ ($1 \leq x \leq 4$; TM = transition metal; X = Si, P, or S), NASICONs offer exceptional thermal stability and compositional flexibility, enabling systematic optimization of redox activity, voltage, and capacity.[30,31]

NASICONs theoretically allow the intercalation of up to four $Na^+$ per formula unit (f.u.), offering the potential for high energy densities. In practice, however, achieving fully reversible (de)intercalation of four $Na^+$ per formula unit remains challenging due to voltage polarization, limited redox accessibility and synthesis constraints. For example, $Na_3V_2(PO_4)_3$ (NVP) can reversibly exchange two $Na^+$ via the $V^{4+}/V^{3+}$ redox couple, operating at an average voltage of ~3.4 V and delivering a practical cathode-level energy density of ~370 Wh/kg.[32] Although insertion of a third $Na^+$ is possible electrochemically (to form $Na_4V^{2+}V^{3+}(PO_4)_3$ at ~1.6 V vs. $Na/Na^+$), the associated large voltage gap between the $V^{4+}/V^{3+}$ and $V^{3+}/V^{2+}$ couples, along with the synthetic difficulty of $Na_4V_2(PO_4)_3$, limits its practical use.[33] A recent study also reported the extraction of the "last" Na in $NaV_2(PO_4)_3$, achieving compositions closer to $V_2(PO_4)_3$ while retaining the NASICON structure, with possible improvements in reversible capacity.[34] Despite these advances, enhancing the overall capacity and energy density of NASICON cathodes remains a key challenge.

To address these limitations, binary and multi-transition-metal (multi-TM) NASICON systems have been investigated as a strategy to activate multiple redox centers and improve electrochemical performance. Compounds such as $Na_3MnTi(PO_4)_3$[35,36] and Fe-Mn-V variants[37–39] demonstrate improved cycling stability, wider voltage windows due to multi-redox activity, and enhanced rate capability compared to unary systems. For instance, $Na_4MnV(PO_4)_3$ delivers high Coulombic efficiency and long-term durability over 1000 cycles.[37] Nevertheless, practical limitations persist, including incomplete utilization of available redox couples and residual structural distortions, such as the Jahn–Teller (J-T) effects in Mn-rich compositions. Additionally, Nb-based NASICONs have demonstrated the ability to achieve higher capacities via the removal of the last Na, albeit at low (de)intercalation voltages, making them more suitable for anode applications.[40] The shortcomings of the unary and binary TM NASICONs highlight the need for systematic multi-TM substitution as a versatile approach to simultaneously activate multiple redox reactions, enhance structural robustness, and enable higher energy densities.

In this study, we employ density functional theory (DFT)-based computations to systematically investigate a series of nine NASICON compositions spanning unary, binary, and ternary systems with Mn, Cr, and Fe substitutions. Specifically, we consider unary systems including $Na_xMn_2(PO_4)_3$ (NMP), $Na_xCr_2(PO_4)_3$ (NCP), $Na_xFe_2(PO_4)_3$ (NFP), binary systems comprising $Na_xMnCr(PO_4)_3$ (NMCP), $Na_xCrFe(PO_4)_3$ (NCFP), $Na_xFeMn(PO_4)_3$ (NFMP) and ternary systems spanning $Na_xMn_{1.0}Cr_{0.5}Fe_{0.5}(PO_4)_3$ (NMCFP), $Na_xCr_{1.0}Fe_{0.5}Mn_{0.5}(PO_4)_3$ (NCFMP), $Na_xFe_{1.0}Mn_{0.5}Cr_{0.5}(PO_4)_3$ (NFMCP). We choose Mn, Cr, and Fe as the TMs in the NASICON framework based on their earth abundance, complementary redox properties, and possible roles in structural stabilization. Through our DFT calculations, we establish systematic trends in the structural features, phase stability and thermodynamic behavior, intercalation voltages across the full Na composition range ($1 \leq x \leq 4$), electronic structure, redox activity, and $Na^+$ migration barriers, and validate with available experimental data, where possible. Notably, we identify NMCFP as a promising ternary NASICON SIB cathode material, offering an optimal balance across thermodynamic stability, phase behavior, intercalation voltages, electronic structure, and $Na^+$ mobility. Our study establishes a comprehensive computational mapping of electrochemical and structural trends across Mn-, Cr-, and Fe-containing NASICONs, thus providing fundamental insights that enable the rational design of high-performance, sustainable NASICON cathodes for next-generation SIBs.

## 2. Methods

We performed all spin-polarized calculations using the Vienna ab initio simulation package.[41,42] We used the projector augmented-wave method[43,44] to describe the core electrons and compiled the potentials used in Table S1 of the supporting information (SI). To account for the strong electronic correlations of the TM *d* electrons, we employed the Hubbard *U* corrected strongly constrained and appropriately normed (i.e., SCAN+*U*) functional.[45,46] We set the effective *U* values as follows: 3.1 eV for Fe, 2.7 eV for Mn, and 0 eV for Cr, based on prior benchmarks (Table S2).[47–49] We used rhombohedral $Na_3V_2(PO_4)_3$ as the starting structure for our calculations, as obtained from the inorganic crystal structure database (ICSD).[50] We truncated the plane-wave basis set at a kinetic energy cutoff of 520 eV, sampled the irreducible Brillouin zone using Γ-centered Monkhorst-Pack[51] *k*-point meshes with a minimum density of 48 *k*-points per Å, and integrated the Fermi surface with Gaussian smearing of width 0.05 eV. We fully relaxed the cell volumes, cell shapes, and ionic positions of all structures considered, without preserving any underlying symmetry, until the residual forces fell below |0.03| eV/Å and the total energies converged within 0.01 meV. We initialized all TMs in high-spin ferromagnetic configurations to approximate their ground-state magnetic ordering. We evaluated the electronic structures by calculating the total and projected densities of states (DOS) with denser 96 *k*-point per Å meshes using the tetrahedron method with Blöchl corrections.[52]

For calculating the average (de)intercalation voltages of Na in the NASICON frameworks considered, we modeled the electrochemical intercalation process as a reversible reaction,

$$Na_yTM_2(PO_4)_3 + (x-y)Na^+ + (x-y)e^- \xleftrightarrow{-\Delta G^0} Na_xTM_2(PO_4)_3 \quad (1)$$

Here, y and x are the charged and discharged Na compositions in the NASICON framework, respectively. We approximated the Gibbs free energy change $\Delta G^0$ using DFT total energies and neglected $pV$ and entropic terms. Subsequently, we calculated the average intercalation voltage as,[53]

$$V = -\frac{\Delta G^0}{(x-y)F} \approx -\frac{E(Na_xTM_2(PO_4)_3) - [E(Na_yTM_2(PO_4)_3) + (x-y)\mu_{Na}]}{(x-y)F} \quad (2)$$

where $\mu_{Na}$ represents the chemical potential of bulk sodium metal in its ground state body centered cubic structure, and F is the Faraday's constant.

We evaluated the phase behavior at 0 K for all intermediate Na-compositions within each NASICON framework considered via DFT-calculated formation energies, $E_f(x)$, with $1 \leq x \leq 4$. For calculating $E_f(x)$, we used the fully charged ($E(Na_1TM_2(PO_4)_3)$) and fully discharged ($E(Na_4TM_2(PO_4)_3)$) compositions as references.

$$E_f(x) = E(Na_xTM_2(PO_4)_3) - \left(\frac{4-x}{3}\right)E(Na_1TM_2(PO_4)_3) - \left(\frac{x-1}{3}\right)E(Na_4TM_2(PO_4)_3) \quad (3)$$

At each intermediate Na composition, we systematically generated symmetrically distinct Na/vacancy orderings using the pymatgen library,[54] in Na compositional steps ($\Delta x$) of 1 across the compositional range $1 \leq x \leq 4$. We used the primitive cell of the NASICON structure, containing 2 f.u., for performing our enumerations. Upon enumeration, we used the Ewald summation technique[55] to rank structures based on electrostatic energies, calculated by assigning point charges of +1, +5, and -2 for the Na, P, and O, respectively, while the TM was assigned an oxidation state that would render the overall structure charge-neutral. Upon electrostatic ranking, we chose the 10 lowest energy structures for further DFT calculations at each Na composition.

For assessing the overall thermodynamic stability of all NASICON compositions considered, we constructed the 0 K convex hull of the Na–Mn–Cr–Fe–P–O six component chemical space using DFT-calculated energies and computed the associated energy above the convex hull ($E_{hull}$) for NASICON compositions considered. For constructing the 6D convex hull, we incorporated all relevant elemental, binary, ternary, quaternary, and quinary ordered structures, as available in the ICSD to ensure a comprehensive stability evaluation.[50] Since only DFT-calculated total energies at 0 K were used to construct the convex hull, the resulting phase diagrams do not explicitly account for entropic or $pV$ contributions.

To estimate the $Na^+$ migration barriers ($E_m$), we adopted a two-step procedure using the fully sodiated $Na_4TM_2(PO_4)_3$ compositions as the starting point for all nine NASICON systems. First, we used the multi atomic cluster expansion (MACE) universal machine learned interatomic potential, namely the MACE-MP-0 potential[56] as integrated within the atomic simulation environment[57] to perform nudged elastic band (NEB) calculations.[58] We relaxed initial and final endpoints with a quasi-Newton optimizer until forces dropped below |0.05| eV/Å over a maximum of 1000 steps. We generated seven intermediate images using the image dependent pair potential interpolation method[59] with a spring constant of 5 eV/Å$^2$, similar to our previous work.[60] We optimized the minimum energy path with the NEB method[61] until band forces converged below |0.05| eV/Å, using the Broyden–Fletcher–Goldfarb–Shanno optimizer.[62] All MACE-based NEB calculations employed the "large" MACE-MP-0 model with maximal message equivariance (i.e., L_max = 2). Second, we re-optimized the MACE-relaxed endpoints using DFT with fixed lattice parameters and relaxing ionic positions. We applied convergence thresholds of 0.01 meV for energy and |0.03| eV/Å for forces. Subsequently, we refined the migration path by combining MACE-relaxed intermediate image

coordinates with DFT-optimized endpoints as inputs for full DFT-NEB calculations. We carried out the DFT-NEB simulations using the Perdew-Burke-Ernzerhof version of the generalized gradient approximation functional,[63] to reduce computational costs while retaining robust qualitative trends.[64]

## 3. Results

### 3.1 Structure

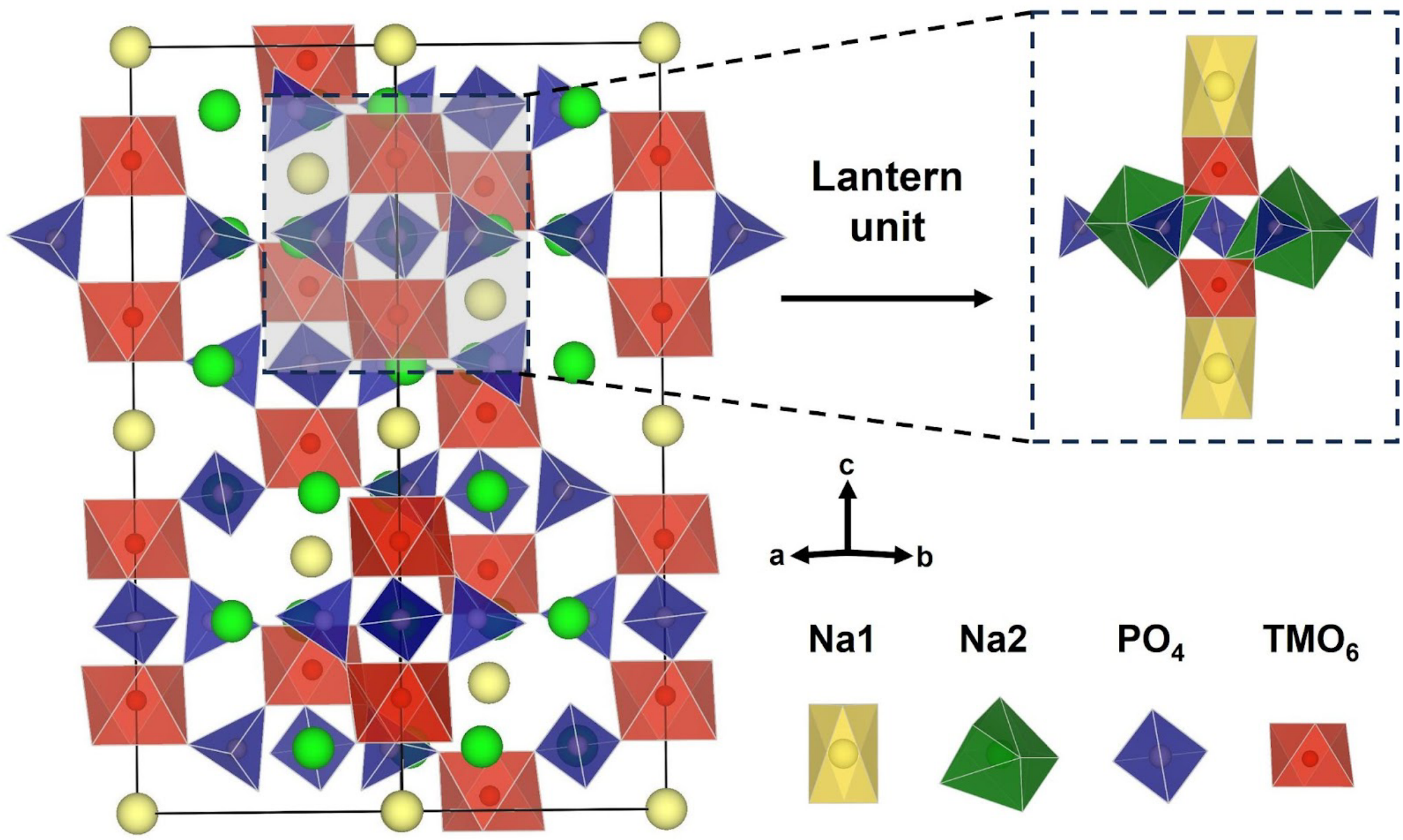

**Figure 1.** Crystal structure of the rhombohedral NASICON $Na_4TM_2(PO_4)_3$ (space group: $R\bar{3}c$) with TM = Mn, Cr, or Fe. The framework is constructed with "lantern units" (highlighted by the dashed box), each consisting of two $TMO_6$ octahedra (red) bridged by three $PO_4$ tetrahedra (blue) via corner-sharing. Sodium occupies two crystallographically distinct sites: Na1 (yellow spheres; six-fold coordinated; 1 per f.u.) located between TM octahedra, and Na2 (green spheres; eight-fold coordinated; 3 per f.u.) coordinated by adjacent phosphate groups.

Generally, NASICON frameworks crystallize in a rhombohedral structure (with $R\bar{3}c$ space group),[65,66] although monoclinic ($C2/c$ or $Cc$)[67,68] and, more rarely, triclinic ($P\bar{1}$)[69] forms have also been reported. The archetypal rhombohedral NASICON structure as displayed in **Figure 1**, comprises corner-sharing $TMO_6$ octahedra and $PO_4$ tetrahedra, arranged into characteristic "lantern units" (inset of **Figure 1**). Each lantern unit consists of two $TMO_6$ octahedra bridged by three $PO_4$ groups, and these motifs assemble into a three-dimensional framework with well-defined $Na^+$ diffusion channels. Na resides in two distinct crystallographic sites: the six-fold coordinated Na1 site located between two TM octahedra, and the eight-fold coordinated Na2 site that lie amidst $PO_4$ groups. At full sodiation (x = 4), one Na1 and all three Na2 sites per f.u. are occupied, and the framework stabilizes in the high-symmetry rhombohedral phase.[70] At x = 1, typically only the Na1 site is occupied and all Na2 sites are vacant.[71] Note that the rhombohedral symmetry of the NASICON structure can be broken, resulting in a monoclinic structure, by the ordering of $Na^+$ and vacancies among the Na2 sites, as typically observed in $Na_xV_2(PO_4)_3$ system at x = 3.[72] Upon DFT relaxation, we observe the optimized lattice parameters of all binary and ternary NASICONs to predominantly preserve the rhombohedral framework with minor distortions, specifically, ~0.07 Å deviations along the *a* and the *b*, ~0.06 Å deviation along the *c* lattice parameters, and angular deviations being <0.4°. Schematics of all ground state Na-vacancy configurations and the associated lattice parameters among binaries and ternaries are compiled in the SI (see Table S3 and Figures S1-S2).

### 3.2 Phase behaviour

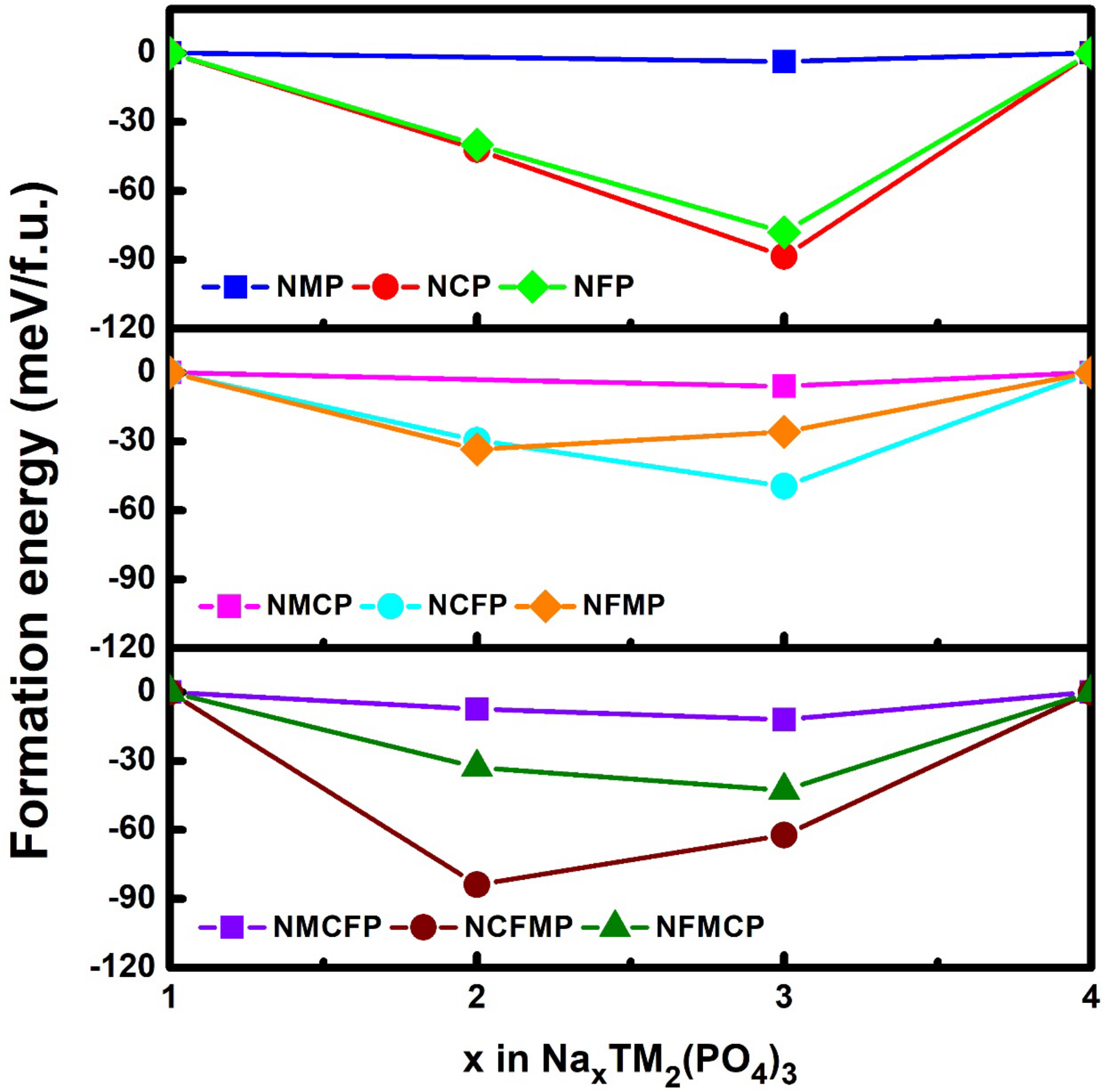

**Figure 2.** DFT-computed $E_f(x)$ and corresponding 0 K convex hulls for Na-vacancy orderings as a function of sodium content (x) in $Na_xTM_2(PO_4)_3$ NASICONs. Each panel represents a compositional class: (a) unary (NMP, NCP, NFP), (b) binary (NMCP, NCFP, NFMP), and (c) ternary (NMCFP, NCFMP, NFMCP). $E_f(x)$ are referenced to the fully charged ($Na_4TM_2(PO_4)_3$) and the fully discharged ($Na_1TM_2(PO_4)_3$) states and are normalized per $M_2(PO_4)_3$ f.u. to enable direct comparison across different compositions. At each sodium content, the $E_f(x)$ of only the most stable Na-vacancy ordering is shown.

We compile the $E_f(x)$ and the resultant pseudo-binary convex hulls at 0 K, for all the NASICON systems considered in **Figure 2**, with the top, middle, and bottom panels consisting of data for the unary, binary, and ternary TM cases. Convex hull displaying the metastable configurations are compiled in Figures S3-S5 for all systems considered. Notably, all convex hulls represent a lack of phase separating behavior upon Na (de)intercalation across the entire range of x considered, given the stability of at least one intermediate Na composition with respect to the end member compositions at x = 1 and x = 4. Also, in all the NASICON compositions considered, the x = 3 phase is always stable with respect to the end-member compositions, which can be attributed to the stability of the 3+ oxidation states of the TMs considered and the specific Na-vacancy ordering that forms at these compositions. Indeed, $Cr^{3+}$ and $Fe^{3+}$ are stable oxidation states for Cr and Fe, respectively, given the half-filled *t2g* and *d* shells of the corresponding cations, which is reflected in the large negative $E_f(x)$ values for Cr-rich and Fe-rich NASICON compositions. On the other hand, $Mn^{3+}$ is J-T active and is not a particularly favored oxidation state for Mn, which is reflected in fairly small negative values in Mn-rich NASICONs. Thus, in Mn-rich compounds, the stability of the x = 3 composition is primarily attributable to the electrostatic stabilization arising from the Na-vacancy ordering among the Na1 and Na2 sites.

Among the unary systems (top panel of **Figure 2**), NCP exhibits the deepest $E_f(x)$ minimum at x = 3, followed by NFP and NMP, indicating qualitatively greater stabilization upon partial (de)sodiation. The steep profile of $E_f(x)$ in NCP suggests a strong thermodynamic preference for phase separation across the x = 1-3 and x = 3-4 compositional ranges. However, the minimum at x = 3 may induce voltage hysteresis upon cycling and increase the difficulty to electrochemically exchange Na across the entire x = 1 to x = 4 range. NFP shows a similar profile to NCP, whereas NMP exhibits a shallow hull with weak intermediate-phase stabilization, possibly indicating the potential to have a solid-solution behavior across the entire x = 1 to x = 4 range. However, the 'shallow' convex hull of NMP can indicate potential structural instability associated with $Mn^{3+}$ J-T distortions.[73,74]

Among the binaries (middle panel of **Figure 2**), NMCP retains a shallow convex hull similar to NMP, implying limited stabilization through Na ordering and the dominant role played by J-T distortions of $Mn^{3+}$. The NCFP profile shows a large degree of similarity with the NCP profile, but with an overall shallower convex hull, indicating the dominant role of Cr-redox in the structure. Interestingly, NFMP shows a ground state configuration at $x = 2$, which is qualitatively different from NFP, indicating the mixed redox-activity of both Fe and Mn in the structure influencing the extent of stabilization at intermediate sodium compositions.

The ternary systems (bottom panel of **Figure 2**) display some interesting trends. NMCFP resembles NMP and NMCP, suggesting that Cr and Fe substitution in a Mn framework does not significantly enhance phase stability, and the electrostatic stabilization offered by Na-vacancy ordering is offset to an extent by the J-T distortions of $Mn^{3+}$. Thus, we do not expect NMCFP to exhibit a significantly strong phase separation behavior compared to unary and binary systems. Similar to NCP and NCFP, NCFMP shows the deepest pseudo-binary convex hull among the ternaries but with the important difference of having the minimum at $x = 2$ instead of $x = 3$. This highlights unique redox and structural interactions, likely dominated by Cr's redox activity, but also modified by the presence of Mn and Fe, similar to thermodynamic trends observed in high-entropy systems.[75,76] NFMCP exhibits a $E_f(x)$ profile that is intermediate to the other ternaries, similar to trends observed in Fe-rich compositions among the unary and binary compositions. Overall, our calculated convex hulls demonstrate how specific TM combinations govern pseudo-binary phase stability across the Na intercalation range which critically affects voltage profiles, hysteresis, and long-term cycling stability in NASICON cathodes.

### 3.3 Intercalation voltage

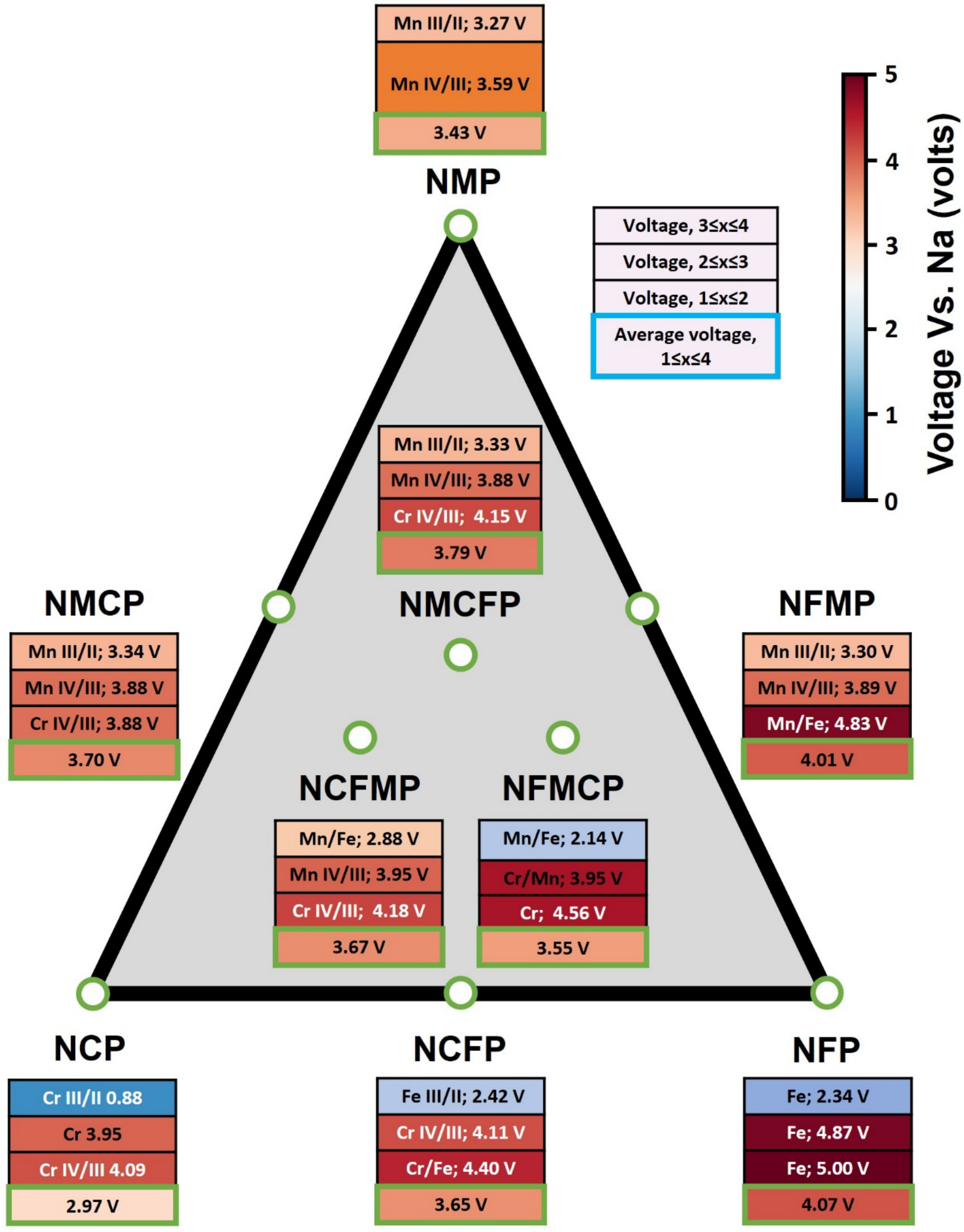

**Figure 3.** DFT-calculated intercalation voltage map of unary (NMP, NCP, NFP), binary (NMCP, NCFP, NFMP), and ternary (NMCFP, NCFMP, NFMCP) NASICON compositions, over the sodium content range $1 \leq x \leq 4$. Each composition is positioned within the ternary compositional triangle, with the vertices, edge midpoints, and interior points representing unaries, binaries, and ternaries, respectively. Each box displays the voltages over specific Na (de)intercalation content ranges, i.e., $3 \leq x \leq 4$, $2 \leq x \leq 3$, and $1 \leq x \leq 2$, with the average (de)intercalation voltage across the $1 \leq x \leq 4$ range highlighted in the green boxes. Voltage values are associated with redox couples, as inferred from computed on-site magnetic moments or known oxidation state trends.

**Figure 3** presents a comprehensive DFT-calculated voltage map (versus $Na/Na^+$) of the $Na_xTM_2(PO_4)_3$ NASICON systems considered across the Mn-Cr-Fe compositional space. The unary NASICONs (NMP, NCP, NFP) form the vertices of the triangle, the binaries (NMCP, NCFP, and NFMP) constitute the edge-centers, and the ternaries (NMCFP, NCFMP, NFMCP) corresponds to the interior circles. Each box indicates the calculated voltage across specific Na concentration ranges that typically correspond to specific TM redox activity, namely, $3 \leq x \leq 4$ (top), $2 \leq x \leq 3$ (middle), and $1 \leq x \leq 2$ (bottom), with the green highlighted bottom box indicating the average voltage across the entire Na composition range ($1 \leq x \leq 4$). Note that the redox activity of the TM is inferred from either the computed on-site magnetic moment changes or based on typical oxidation state trends that are observed.

The calculated average voltage spans from a low 2.97 V (NCP) to a high 4.07 V (NFP) vs. $Na/Na^+$, with other NASICON compositions exhibiting average voltages that are intermediate to these extreme values. Fe-rich compositions (NFP, NFMP) consistently deliver the highest average voltages (~4.01-4.07 V), reflecting the high voltages at which $Fe^{4+}/Fe^{3+}$ redox occurs. Cr-rich compositions, such as NCP and NCFP show lower average voltages (2.97 V and 3.65 V), as governed primarily by $Cr^{3+}/Cr^{2+}$ and $Cr^{4+}/Cr^{3+}$ redox activities,[77] while Mn-based compositions exhibit intermediate voltages (3.43 V for NMP and 3.70 V for NMCP), involving mainly $Mn^{3+}/Mn^{2+}$ and $Mn^{4+}/Mn^{3+}$ transitions. Given that the Fe-rich NASICONs (NFP, NFMP, and NFMCP) exhibit voltage plateaus that occur at voltages higher than 4.5 V vs. Na at low levels of Na, which may be beyond the electrolyte stability windows of Na-based electrolytes,[78] we do not expect large degrees of electrochemical Na deintercalation to be viable in Fe-rich unary and binary NASICONs considered.

The trends in average voltages among ternaries are different from the trends observed among unaries and binaries. For example, the Fe-rich ternary (NFMCP) exhibits the lowest average voltage (~3.55 V), which is dominated by the $Mn^{4+}/Mn^{3+}$ and $Cr^{4+}/Cr^{3+}$ redox couples instead of the high-voltage $Fe^{4+}/Fe^{3+}$ couple, with the low-voltage $Fe^{3+}/Fe^{2+}$ redox couple participating at high degrees of sodiation in NFMCP. The Mn-rich NMCFP yields the highest average voltage (~3.79 V), with the redox transitions comprising exclusively of Mn/Cr and Fe largely being inert. The Cr-rich NCFMP exhibits an intermediate average voltage (~3.67 V), which is also dominated by Mn and Cr redox transitions, with Fe participating at high concentrations of Na. Notably, the NMCFP NASICON yields the highest average voltage among all systems considered where the individual voltage plateaus (from $1 \leq x \leq 2$ to $3 \leq x \leq 4$) lie below the oxidative stability limit of most Na electrolytes (~4.5 V vs. Na)[78], thus representing the most promising NASICON cathode among the compositions considered, in terms of average voltages.

In general, redox transitions in multi-TM NASICON systems display remarkable consistencies, i.e., a given redox couple occurs over a small voltage range, with electrostatics and local chemical environments playing a minor role.[74] Indeed, we do observe redox couples occurring at similar voltages in our NASICON systems as well. For example, the $Mn^{3+}/Mn^{2+}$ and $Mn^{4+}/Mn^{3+}$ redox couple occur within narrow 3.30-3.34 V and 3.88-3.95 V ranges, respectively, across binary and ternary NASICONs containing Mn. Similarly, $Cr^{4+}/Cr^{3+}$ redox occurs across a 4.09-4.18 V window across all Cr-containing NASICONs, except NMCP (at 3.88 V). The exceptions to redox couples occurring over narrow intercalation voltages predominantly occur in Fe-containing systems, which can be attributed to a lack of clear signature of Fe redox couples (especially associated with the $Fe^{4+}/Fe^{3+}$ redox) and possible contributions from the anion in the redox process itself. Thus, our work highlights the trends, the role of TM redox couples, and other factors such as electrostatics that influence the voltages at which Na can (de)intercalate in the multi-TM NASICON frameworks.

### 3.4 Electronic structure

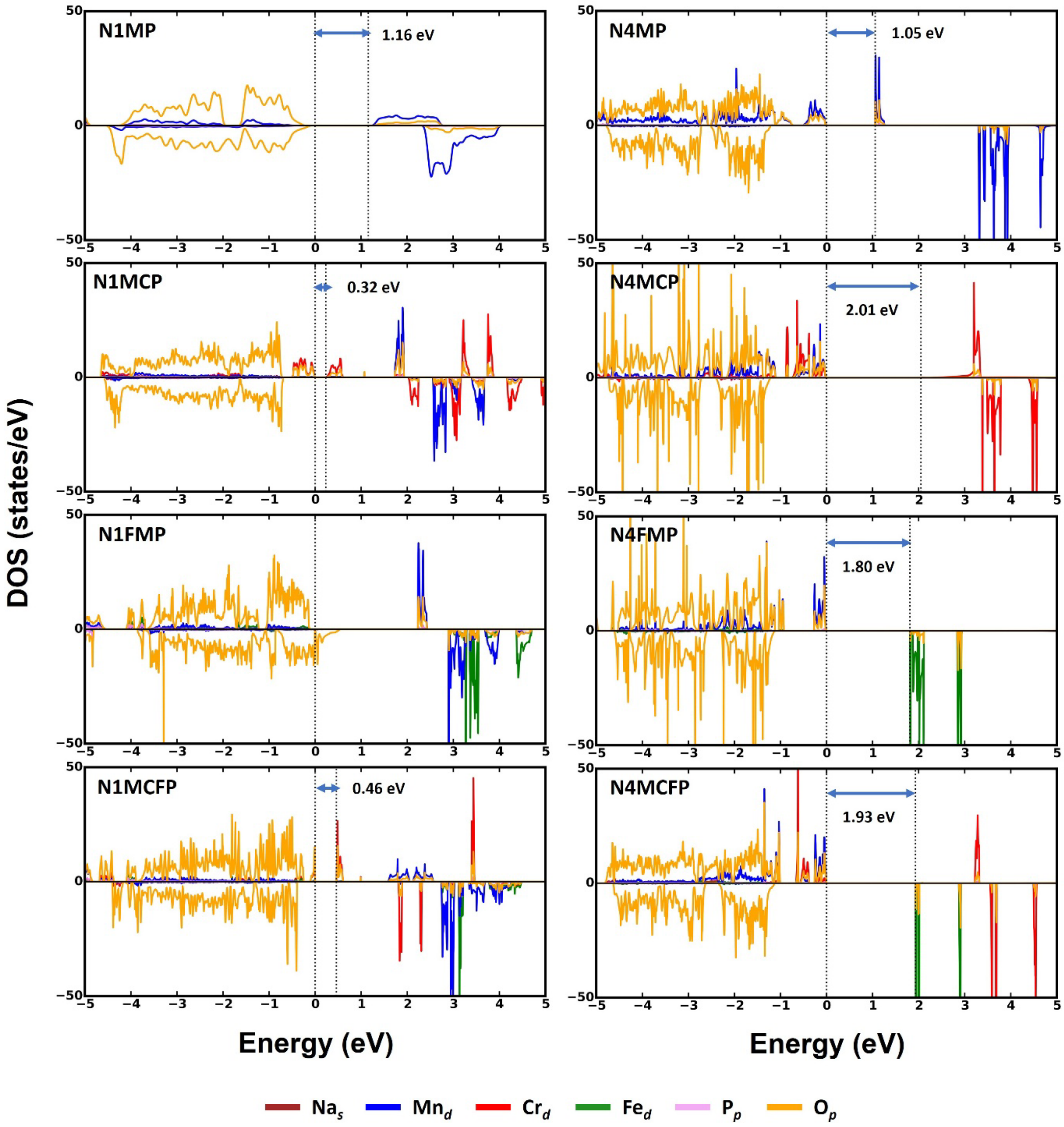

**Figure 4.** Atom-projected electronic DOS for select $Na_xTM_2(PO_4)_3$ NASICON systems, where TM = Mn, Cr, and/or Fe. Panels on the left and right columns correspond to charged (x = 1) and discharged (x = 4) Na-compositions, respectively. We compile the DOS for four NASICONs, namely, NMP (top row), NMCP (second), NFMP (third), and NMCFP (bottom row). The projected DOS contributions from Na *s*, Mn *d*, Cr *d*, Fe *d*, P *p*, and O *p* orbitals are shown using distinct colors, as indicated in the legend. The vertical dotted lines mark the valence

(set to 0 eV) and conduction band edges in non-metallic systems, and the Fermi energy (set to 0 eV) in metallic systems. For non-metallic systems, the numerical annotations correspond to the calculated $E_g$.

**Figure 4** presents the atom-projected DOS for $Na_xTM_2(PO_4)_3$ NASICON systems in the fully charged (x = 1, left column) and fully discharged (x = 4, right column) states across four representative compositions, namely, NMP, NMCP, NFMP, and NMCFP. We plot the DOS over an energy range of –5 eV to +5 eV relative to either the Fermi level (set at 0 eV in metallic systems) or the valence band edge (set to 0 eV in non-metallic systems), with the spin-up and spin-down states shown as positive and negative DOS values, respectively. The number adjacent to Na in the NASICON notation (for example 4 in 'N4MCP') indicates that x = 4 within the NASICON (i.e., the composition is $Na_4MnCr(PO_4)_3$). The DOS of the all NASICON compositions considered are compiled in the SI, see Figures S6-S14.

Across all compositions in **Figure 4**, the valence band edge or the Fermi level predominantly comprises O 2*p* states, with the TM 3*d* states contributing significantly in several systems, such as N4MP, N1CP, N4CP, N4FMP, and N4MCFP, similar to observations in other NASICON compositions.[74] The conduction band mainly consists of unoccupied TM *d* states, which is typically consistent with their redox activity during sodium (de)intercalation.[79] In general, as sodium content increases (from charged to discharged state), the TM *d* states shift toward the valence band/Fermi level, progressively filling the electronic states, and thus narrowing the band gap ($E_g$) in several unaries.[80] For example, the $E_g$ in NMP reduces from the charged (N1MP, ~1.16 eV) to the discharged (N4MP, ~1.05 eV) along with a shift of Mn *d* states from the conduction band to the valence band. Note that phase transitions within the unary NASICONs, such as the rhombohedral to monoclinic transition in $Na_xV_2(PO_4)_3$ at x = 3, can cause non-monotonic variations in the band gap with sodiation.[72]

However, such progressive filling of TM *d* states does not cause a drop in $E_g$ in the binary and ternary systems. For instance, the charged state in NMCP exhibits a low $E_g$ (~0.32 eV), which can be attributed to sharp, localized Cr *d* at both the valence and conduction band edges (i.e., the $E_g$ falls among the Cr *d* bands). Upon full sodiation, the $E_g$ widens to ~2.01 eV, with the DOS indicating the valence and conduction band edges being dominated by Mn *d* and Cr *d* states, respectively (i.e., the $E_g$ falls between the Mn and Cr *d* states). The shift of the Mn and Cr *d* states with increasing Na content in NMCP is also consistent with the sequence of redox couples that are activated (see **Figure 3**) with Na insertion, i.e., the $Cr^{4+}/Cr^{3+}$ redox couple activates first (leading to a shift in the Cr *d* states to the valence band) followed by $Mn^{4+}/Mn^{3+}$ and $Mn^{3+}/Mn^{2+}$ couples (resulting in the shift of Mn *d* states). We observe similar increases in $E_g$ in NFMP (0 to 1.80 eV) and NMCFP (0.46 to 1.93 eV) systems, which are accompanied by a change in the *d*/*p* states surrounding the Fermi level with increasing Na content that are consistent with the sequence of redox reactions occurring in these systems. Thus, the presence of multiple TMs may not reduce the $E_g$ of NASICON systems with increasing sodiation, at least at the fully discharged compositions, and hence may not necessarily contribute to any improvement in electronic conductivity with sodiation.

### 3.5 Thermodynamic stability

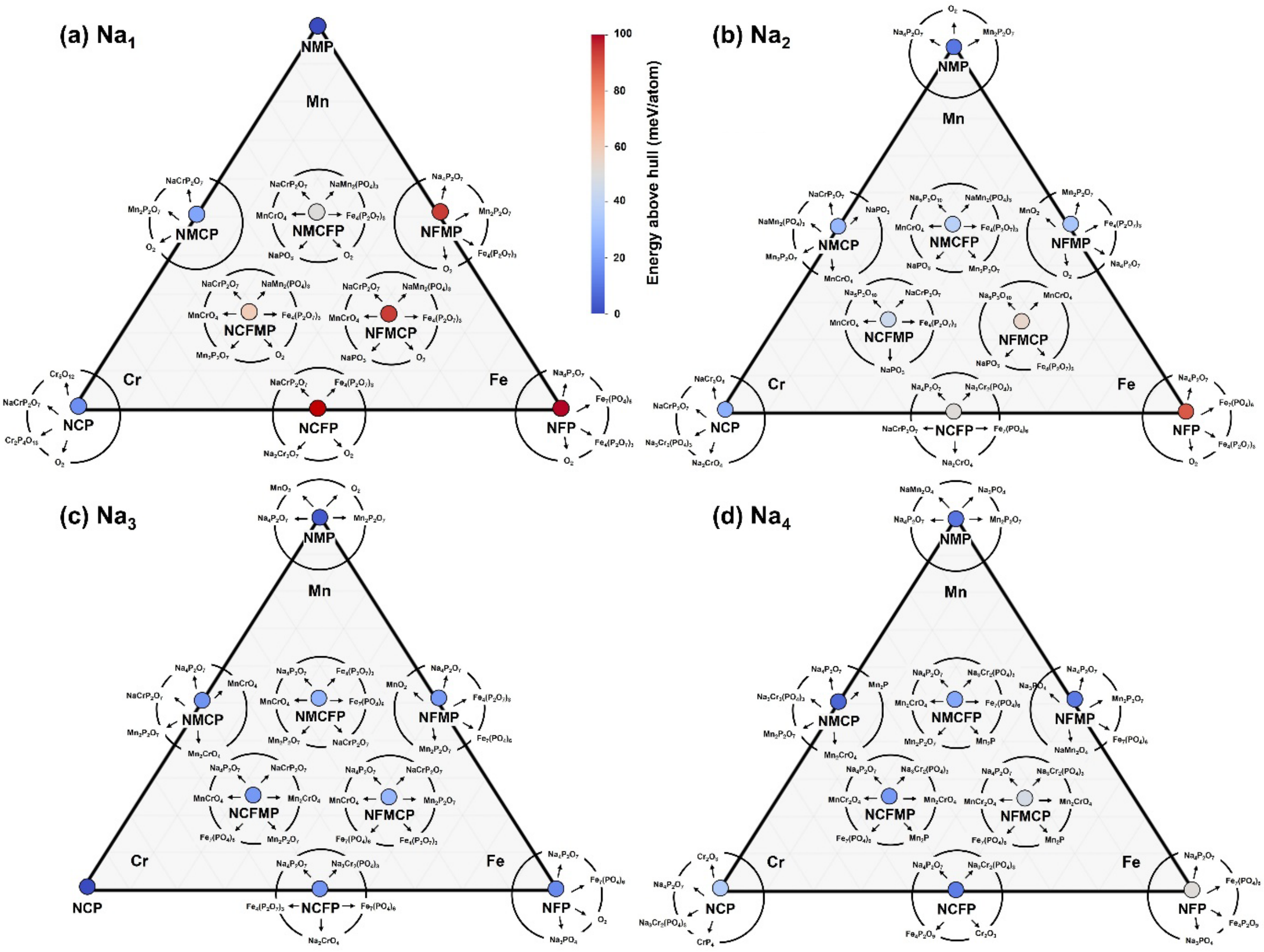

**Figure 5.** The stability of the NASICON systems considered, as calculated based on the 0 K Na-Mn-Cr-Fe-P-O 6D convex hull, are visualised as ternary projections for different Na compositions of $Na_xTM_2(PO_4)_3$, namely, (a) x =1, (b) x = 2, (c) x = 3, and (d) x = 4. Stable ($E_{hull}$ = 0 meV/atom), metastable ($E_{hull} \leq 50$ meV/atom), and unstable ($E_{hull} > 50$ meV/atom) NASICONs are displayed as dark blue, blue to orange, and red circles, respectively. Arrows surrounding metastable and unstable compositions indicate the stable decomposition products at the corresponding compositions.

To quantify the thermodynamic stability of the various NASICON systems considered, we calculate the $E_{hull}$ based on the Na-Mn-Cr-Fe-P-O 6D system and visualize the data as ternary projections at different Na compositions in **Figure 5**. Specifically, we consider the unary NASICONs, namely, NMP, NCP, and NFP, as end compositions and plot the $E_{hull}$ values for Na contents of x = 1 (panel a), x = 2 (panel b), x = 3 (panel c), and x = 4 (panel d). Note that all stable phases lie on the hull ($E_{hull}$ = 0, dark blue circles in **Figure 5**), metastable phases are considered to exhibit an $E_{hull}$ within a threshold of 50 meV/atom (blue to orange circles), and unstable phases exhibit $E_{hull} > 50$ meV/atom (red circles). The $E_{hull}$ threshold of 50 meV/atom is a rule-of-thumb and reflects the fact that compounds slightly above the convex hull can often be realized experimentally under high temperature or non-equilibrium synthesis conditions.[81–83] For all metastable and unstable NASICONs in **Figure 5**, the arrows and associated chemical formula indicate the stable decomposition products that the NASICONs are expected to form under equilibrium. A heatmap summarizing all $E_{hull}$ values across all compositions is compiled in Figure S15.

The unary NMP and NCP remain stable or metastable ($E_{hull} \leq 25$ meV/atom) across the entire Na composition range ($1 \leq x \leq 4$), with the x = 1 and x = 3 compositions being thermodynamically stable, consistent with experimental reports of synthesis and Na-cycling across these compositions.[84] In contrast, NFP exhibits high instability, particularly at low Na content (x = 1 and 2), with $E_{hull}$ exceeding 100 meV/atom at x = 1. The (in)stability of unary NASICONs can largely be attributed to the stability of the underlying TM in its oxidation state. For example, Mn is known to form stable compounds in its +4 oxidation state whereas Fe's +4 oxidation state is known to be unfavorable.[85] Interestingly, our calculations indicate $Na_3Fe_2(PO_4)_3$, a NASICON that has been synthesized and electrochemically cycled, to be metastable ($E_{hull}$ = 14.6 meV/atom), highlighting the fact that metastable compounds can indeed be synthesized and used under room temperature conditions.

Similar to unary systems, binaries show better (meta)stability at higher Na contents (x = 3–4), with $E_{hull}$ values in the range of ~6–20 meV/atom, indicating potential synthesizability and electrochemical use. Indeed, Mn-Cr,[86] Cr-Fe,[87] Fe-Mn[88] binary NASICONs have been experimentally synthesized and electrochemically cycled, in agreement with our data. While NMCP continues to be metastable at lower Na compositions ($E_{hull}$ 22-28 meV/atom), both Fe-containing binaries (NFMP and NCFP) become unstable at x = 1 ($E_{hull}$ > 90 meV/atom), suggesting the continued role of the unfavorable +4 oxidation state of Fe in determining the thermodynamic (in)stability of the NASICON. Thus, among binary NASIONs considered, NMCP is the only candidate favorable for reversible Na (de)intercalation across the entire Na composition range.

Ternary systems generally display higher $E_{hull}$ values across all Na compositions, compared to their binary and unary counterparts, often approaching or exceeding the metastable limit especially at low Na content. For example, at x = 3, the ternaries exhibit $E_{hull}$ in the range of 18.6-29.1 meV/atom, while binaries and unaries exhibit $E_{hull}$ values in the ranges of 15.9-19.1 meV/atom and 0-14.6 meV/atom, respectively. Thus, experimental synthesis and (meta)stability of the ternaries under electrochemical conditions may have to arise from possible configurational entropic contributions due to the presence of multiple types of TMs. For example, the ternary NMCFP is predicted to decompose to $NaMn_2(PO_4)_3$ and $Na_3Cr_2(PO_4)_3$ at low and high Na contents, respectively, indicating that any experimental synthesis needs to avoid the formation of unary (or binary) NASICONs that can act as thermodynamic sinks.

While NMCFP and NCFMP ternaries approach or exceed the metastability threshold at x = 1, the Fe-rich NFMCP ternary exceeds the threshold at x = 2 and reaches a high $E_{hull}$ value of 93.6 meV/atom at x = 1, highlighting the continued contribution of the $Fe^{4+}$ state in causing instability even in higher-component NASICONs. Nevertheless, the overall metastable nature of NMCFP ($E_{hull}$ 23-50.4 meV/atom), the lack of a strong tendency for phase separating behavior with Na (de)intercalation (**Figure 2**), and the high average Na (de)intercalation voltage exhibited by NMCFP (3.79 V; **Figure 3**) indicates the promising nature of NMCFP as a high-performance cathode for SIBs among the NASICONs considered in this work. Notably, NMCFP exhibits its lowest $E_{hull}$ of 23 meV/atom at x = 4 with possible decomposition to the unary $Na_3Cr_2(PO_4)_3$, while it exhibits a slightly higher $E_{hull}$ of 26.3 meV/atom at x = 3 without the formation of any NASICON decomposition products, indicating that x = 3 is a better Na content to target in the experimental synthesis of the NMCFP ternary NASICON than x = 4.

### 3.6 $Na^+$ migration pathways and barriers

Na migration in a given NASICON framework can occur via three principal channels: Na1–Na1, Na2–Na2, and Na1–Na2. Previous studies[89] have indicated that Na1–Na2 hops exhibit the lowest $E_m$ (typically in the 0.25–0.35 eV range) due to favorable migration geometry and minimal lattice strain, whereas direct Na1–Na1 and Na2–Na2 hops often exceed 0.6–1.0 eV due to structural bottlenecks. Hence, we exclusively focus on the symmetrically distinct Na1–Na2 hops across all NASICON compositions considered, at the discharged Na content (x = 4), and compiled the DFT-NEB calculated $E_m$ in **Figure 6**. Note that we did not consider intermediate or charged Na content in the NASICON compositions primarily to reduce computational costs, while the calculated data at x = 4 should provide qualitative guidance and trends across all NASICON systems considered.

Blue, orange, and green symbols in **Figure 6** correspond to unary, binary, and ternary NASICONs considered respectively. In binary and ternary systems, due to the arrangement of different TM ions in the NASICON structure, the number of distinct Na1-Na2 hops increases in comparison to unary system. For example, in NMCP, we identify $Na1_{(1)}$–$Na2_{(3)}$ and $Na1_{(1)}$–$Na2_{(6)}$ as distinct migration channels, validated via pymatgen's StructureMatcher module,[54] where the numerical label in bracket indicates the specific index of the Na site within the structure. Thus, the difference between the solid circles and hollow diamonds among binary and ternary NASICONs in **Figure 6** reflect the range of $E_m$ for the possible and distinct Na1-Na2 hops within the NASICON structure. The numerical annotations in **Figure 6** correspond to the $E_m$ values (in meV), with both the lower and upper limiting values of $E_m$ indicated for the binary and ternary systems. The minimum energy pathway profiles associated with all DFT-NEB $E_m$ calculations are compiled in Figure S16-S18.

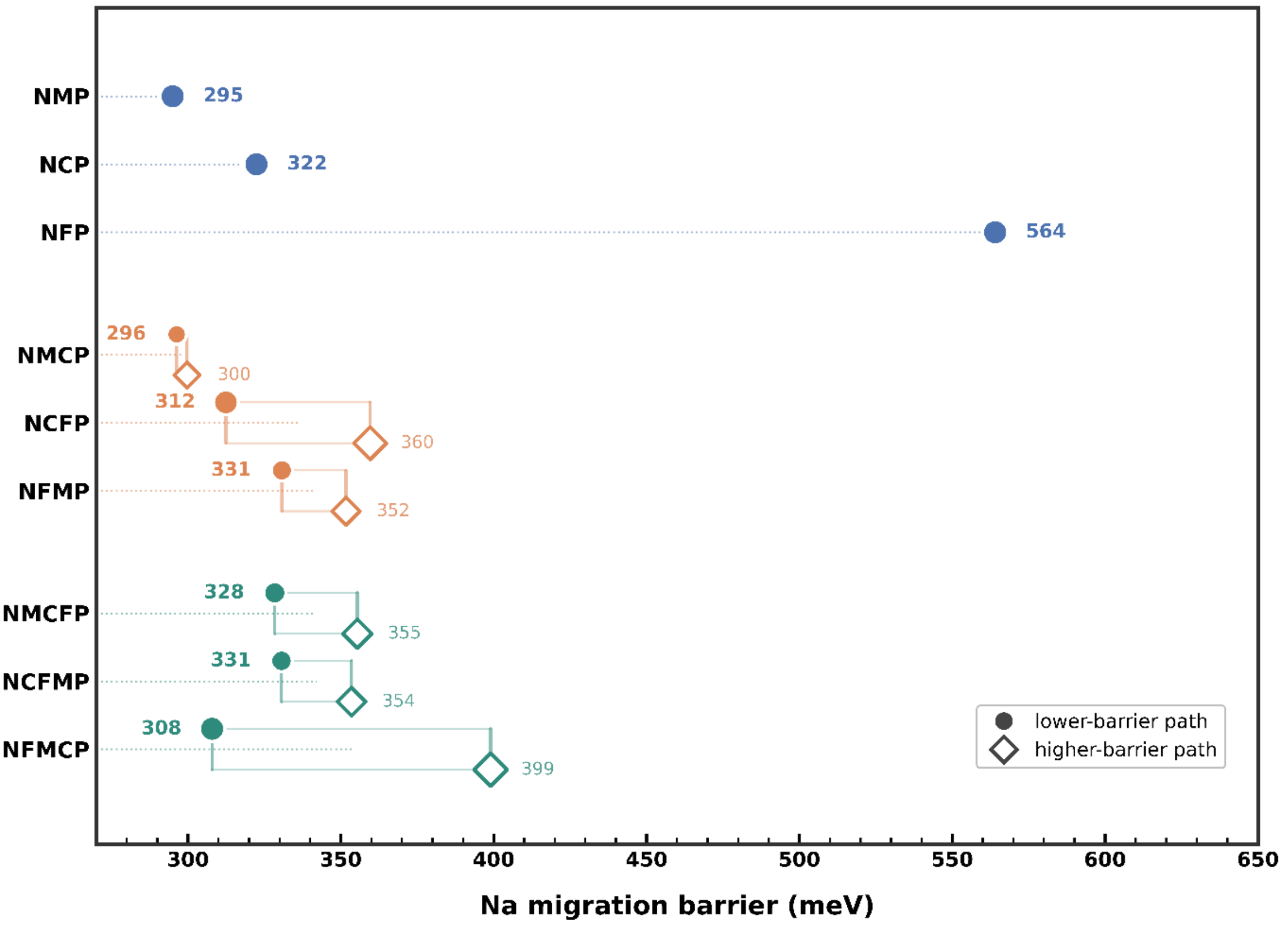

**Figure 6.** Calculated $Na^+$ $E_m$ for unary (blue symbols), binary (orange), and ternary (green) NASICON systems. For binaries and ternaries, the solid circles and hollow diamonds indicate the lower and upper limits, respectively of symmetrically distinct Na1–Na2 hop pairs identified in each structure.

Among the unary NASICONs, NMP and NCP exhibit low $E_m$ of 295 meV and 322 meV, respectively, consistent with fast $Na^+$ transport in these cathodes. Interestingly, NFP showed a much higher $E_m$ of 564 meV, likely arising from lattice distortions and electrostatic pinning of $Na^+$ in the structure, in line with the poor ionic mobility often reported for Fe-based NASICONs.[90] Note that a $E_m$ of 564 meV does not preclude a material from being used as a cathode, with the range of 525-650 meV often used in identifying cathode materials with good ionic mobility.[91] Thus, all the NASICONs considered in this study exhibit $E_m$ lower than 650 meV, indicating that all compositions can be considered for SIB cathode application, in terms of ionic mobility.

All binary NASICONs considered display low-to-moderate $E_m$ values, highlighting swift motion of $Na^+$ within these structures irrespective of the specific TM combination present. Indeed, NMCP, NCFP, and NFMP show $E_m$ values in the ranges of 296-300 meV, 312-360 meV, and 331-352 meV, respectively. Both Fe-containing binary NASICONs, namely NCFP and NFMP exhibit $E_m$ that are significantly below the $E_m$ of NFP (564 meV), highlighting the reduced influence of any $Fe^{3+}$-based pinning effect on the $Na^+$ motion, and underscoring the beneficial role of mixed TM environments in reducing $E_m$.

Ternary NASICONs display broader $E_m$ distributions, compared to the binary systems, reflecting more complex interplay of multiple TMs occupying distinct sites and larger sensitivity of $Na^+$ transport to local TM arrangements. For example, NFMCP exhibits a $E_m$ range of 91 meV (with low and high values of 308 and 399 meV), which is higher than the 21 meV and 48 meV range shown by the NFMP and NCFP, respectively. NMCFP also exhibits a wider range of 27 meV (328-355 meV) compared to NMCP (4 meV) and NFMP (21 meV). The exception to the trend of ternaries exhibiting larger $E_m$ ranges than corresponding binaries is NCFMP shows a $E_m$ range of 23 meV (331-354 meV), which is lower than the 48 meV range shown by NCFP. Nevertheless, all ternaries exhibit both high and low $E_m$ values that are within the 650 meV threshold, suggesting that all ternaries should display sufficiently facile $Na^+$ motion. Importantly, we find the ternary NMCFP NASICON, which was identified as promising based on phase behavior, intercalation voltage, and thermodynamic stability, to be promising in terms of $Na^+$ $E_m$ as well.

## 4. Discussion

Our study systematically explores the structural, thermodynamic, electronic, and transport properties of nine Na-based Mn–Cr–Fe NASICON frameworks (**Figure 1**) across unary, binary, and ternary compositions using DFT-based computations. We identify clear Na and TM composition-dependent trends linking TM chemistry to phase behavior (**Figure 2**), intercalation voltage (**Figure 3**), electronic structure and $E_g$ (**Figure 4**), 0 K thermodynamic stability (**Figure 5**), and $Na^+$ $E_m$ (**Figure 6**). In particular, we observe Cr- and Fe-rich systems to provide stronger thermodynamic stabilization, Mn-containing systems to show lower $E_m$ but exhibit structural distortions, and multi-TM systems introduce competing effects that modify redox activity and $Na^+$ transport. Additionally, unaries exhibit well-defined minima in their 0 K convex hulls and the corresponding redox-driven voltage steps, while binaries and ternaries introduce mixed-TM stabilization and complex cooperative effects that alter phase behavior. Importantly, we identify the ternary NMCFP NASICON to be promising as a SIB cathode based on our computed phase behavior, voltage, stability, and ionic mobility. Our work highlights both the opportunities and challenges of designing high-performance multi-TM NASICON cathodes for SIBs.

One of the approximations involved in our work is the use of a hybrid MACE-MP-0 and DFT-based NEB workflow to calculate to calculate $Na^+$ $E_m$ that we adopted to balance accuracy and computational efficiency. While this strategy captures trends in $E_m$ values, results may vary if alternative universal machine learned interatomic potentials,[60,92] different interpolation schemes,[93] or direct full-DFT NEB calculations are employed. Additionally, we have only calculated $E_m$ values at the discharged composition ($x = 4$, **Figure 6**), with several NASICONs exhibiting non-monotonic trends in $E_m$ values with Na-content and associated bottlenecks for Na (de)intercalation at low Na contents ($x \sim 1$).[94,95] Nevertheless, given that multi-TM substitution has mostly a marginal effect on the calculated $E_m$ values, we expect qualitative trends at lower Na-contents to be similar among the NASICON chemistries considered.

Comparisons with experimental studies provide context to our calculated results. For instance, our high $E_m$ for Fe-based NASICONs (**Figure 6**) aligns with the poor ionic conductivity observed experimentally for $Na_3Fe_2(PO_4)_3$.[96,90] Similarly, the predicted low $E_m$ for Mn- and Cr-rich (unary) systems are consistent with their relatively higher conductivities.[96] However, our convex-hull results suggest that ternary Mn–Cr–Fe NASICONs are metastable ($E_{hull}$ > 50 meV/atom, **Figure 5**), in partial alignment with experimental reports of synthesis in V-containing ternary NASICONs (such as V-Mn-Fe[37–39]). This discrepancy may arise from kinetic stabilization, non-equilibrium synthesis routes, or entropic contributions not captured in our 0 K analysis. Voltage predictions (**Figure 3**) also agree qualitatively with reported redox activity ($Fe^{3+}/Fe^{2+}$ and $Cr^{3+}/Cr^{2+}$ at low voltages, $Mn^{3+}/Mn^{2+}$ at intermediate, $Cr^{4+}/Cr^{3+}$ at high), though the exact voltage values are not always in agreement, possibly due to the approximations involved with DFT-based 0 K calculations.

Some of the observed trends in our work are non-intuitive. For example, unary and binary NASICONs show the global minimum with Na content (**Figure 2)** at x = 3, which is consistent with favorable Na/vacancy ordering within the Na sublattice and stabilization of $TM^{3+}$ oxidation states, often associated with symmetry lowering from rhombohedral to monoclinic structures.[72] On the other hand, the global minimum with Na content in ternaries shifts toward x = 2. One possible explanation is that mixed-TM compositions redistribute redox activity across multiple TM species, altering the relative stability of Na-vacancy configurations at intermediate compositions due to competing local environments and electrostatic interactions. Similarly, the electronic structure results (**Figure 4**) show TM composition-dependent evolution, namely, Mn- and Cr-based unaries exhibit a gradual shift of TM *d* states toward the valence band with sodiation, while multi-TM systems display a redistribution of *d*-state contributions across different TM species, leading to non-monotonic $E_g$ behavior. These trends suggest that orbital hybridization, charge localization, and local structural asymmetry play important roles and warrant further investigation to decouple their influences.

Further extension of our work can incorporate finite-temperature effects via phonon calculations or via constructing cluster expansion Hamiltonians to capture configurational entropy effects, which may be important to determine TM mixing behavior.[97,72] More accurate electronic-structure methods, such as hybrid functionals[80,98] or many-body GW calculations can yield better predictions of $E_g$.[99] On modelling the Na transport, ab initio molecular dynamics simulations[100] can be used to include effects of ionic correlation, which can aid in assessing $Na^+$ diffusivities directly. Finally, more systematic experimental data, on phase behavior, voltages, and Na-mobilities, will aid in further refining of calculations and targeting chemistries that have the most promise.

**5. Conclusion**

We systematically investigated nine NASICON compositions containing earth-abundant Mn, Cr, and/or Fe using DFT calculations as potential SIB cathodes. Specifically, we examined their structural stability, phase behavior, intercalation voltages, electronic structures, and $Na^+$ $E_m$ across unary, binary, and ternary systems. In terms of phase behavior, all systems exhibit stable intermediate compositions, with a strong thermodynamic preference for x = 3 among unaries, while NFMP and NCFMP show a shift in stability toward x = 2, indicating altered Na/vacancy ordering alongside redistribution of redox activity. Fe-rich systems (e.g., NFP, NFMP) deliver the highest average voltages (~4.0 V, due to $Fe^{4+}/Fe^{3+}$ redox), Cr-rich systems show the lowest (~3.0 V), and Mn-containing systems provide intermediate voltages, with multi-TM compositions providing an additional handle to tune voltages. Multi-TM substitution also leads to redistribution of TM *d* states and non-trivial $E_g$ evolution rather than simple band filling as observed in unaries. In terms of thermodynamic stability, all unaries and several binaries remain stable or metastable (0–50 meV/atom), while ternaries mostly lie in the metastable regime particularly at low Na content. With respect to $Na^+$ mobility, Mn- and Cr unaries exhibit low $E_m$ (~0.3 eV), while Fe-unary shows significantly higher $E_m$ (>0.5 eV). Mixed-TM systems do exhibit $Na^+$ $E_m$ that are in a similar vicinity as the NMP and NCP systems, indicating possibly facile $Na^+$ transport. Overall, we identify the ternary NMCFP as a viable candidate offering a balance between voltage, stability, phase behavior, and mobility. Our work provides detailed insights into the role of compositional complexity, phase behavior, and $Na^+$ transport in NASICON frameworks, and offers guidance for the rational design of energy-dense high-performance cathodes for SIBs.

**Author contributions**

S.B. performed all DFT calculations and drafted the initial manuscript. A.K.B. carried out all MACE-based NEB calculations. S.B. and A.K.B. revised and edited the manuscript. S.G.G. supervised all aspects of the work and guided the interpretation of results. All authors approved the final version of the manuscript.

**Acknowledgments**

S.B. acknowledges the Institute of Eminence Postdoctoral Fellowship awarded by the Indian Institute of Science for financial support. G.S.G. acknowledges financial support from the Science and Engineering Research Board (SERB) of the Department of Science and Technology, Government of India, under the sanction number IPA/2021/000007. A portion of the density functional theory calculations reported in this work were carried out using computational resources provided by the Supercomputer Education and Research Center (SERC), Indian Institute of Science. We also acknowledge the National Supercomputing Mission (NSM) for providing access to the 'PARAM Siddhi-AI' and 'PARAM Utkarsh-AI' systems under the National PARAM Supercomputing Facility (NPSF), C-DAC Pune, supported by the Ministry of Electronics and Information Technology (MeitY) and the Department of Science and Technology (DST), Government of India.

**Conflict of interest**

The authors declare no conflict of interest.

**Data and code availability**

All computed data, relevant scripts, and developed codes associated with this study are freely available via our GitHub repository.